\documentclass[final,5p,times,twocolumn,authoryear]{elsarticle}

\usepackage{amssymb}
\usepackage{amsmath}
\usepackage{hyperref}
\usepackage{xcolor}
\usepackage{listings}
\usepackage{orcidlink}

\graphicspath{{figures/}}

\newcommand{\Msun}{M_\odot}
\newcommand{\foraois}{\texttt{foraois}\,}

\journal{Astronomy \& Computing}

\begin{document}

\begin{frontmatter}

\title{\texttt{foraois}: Fast, Flexible Extended Press--Schechter Merger Trees with a General First-Crossing Solver}

\author[first,cor1]{C. Power\orcidlink{0000-0002-4003-0904}}
\cortext[cor1]{Corresponding author}
\ead{chris.power@uwa.edu.au}
\affiliation[first]{
    organization={International Centre for Radio Astronomy Research, The University of Western Australia},
    addressline={35 Stirling Highway},
    city={Crawley},
    postcode={6009},
    state={Western Australia},
    country={Australia}
}

\author[second]{S. Balu\orcidlink{0000-0002-5281-5151}}
\affiliation[second]{
    organization={Facultad de Físicas, Multidisciplinary Unit for Energy Science, Universidad de Sevilla},
    city={Seville},
    postcode={41012},
    country={Spain}
}

\begin{abstract}
We present \foraois (Gaelic for \textit{Forest}; \textsc{FAW-reesh}), an open-source Python framework for generating Monte Carlo dark matter halo merger tree histories. \foraois separates the description of the linear density field, the physics of halo collapse, and the algorithmic construction of halo merger histories, allowing different cosmologies, collapse prescriptions, and tree building approaches to be combined within a common framework. The package implements two approaches to the same merger-tree problem of sampling progenitor masses in the excursion-set picture: the semi-empirical algorithm of \citet{ParkinsonColeHelly2008}, which applies a fitted correction to the EPS progenitor distribution, and the numerical solution of the Markovian first-crossing equation developed by \citet{ZhangHui2006}, which for the default real-space top-hat window is an approximation to the correlated-walk problem.
Linear matter power spectra for Cold, Warm, and Fuzzy Dark Matter (CDM, WDM, and FDM) can be supplied, while the modular collapse treatment provides an extension point for alternative barrier models. The framework also supports constrained merger histories based on the Brownian-bridge construction of \citet{NadlerEtAl2023}. 
We run extensive validation and scaling tests, and also show an illustrative example of coupling \foraois to a light-weight semi-analytical model of galaxy formation.
\end{abstract}


\end{frontmatter}


\section{Introduction}
\label{introduction}

The formation and hierarchical assembly of dark matter haloes underpin theoretical models of galaxy formation and large-scale structure. While halo growth is intrinsically nonlinear and cosmological $N$-body simulations are the standard tool to study it \citep[e.g.][]{AnguloHahn2022}, the extended Press--Schechter (EPS) formalism provides a convenient alternative that allows for a statistical description of halo abundances and assembly histories \citep{PressSchechter1974,BondEtAl1991,Bower1991,LaceyCole1993}. In this excursion set formulation, we calculate a smoothed linear density contrast, $\delta$, that traces a trajectory $\delta(S)$ as the smoothing scale $S$ is varied, where $S(M)=\sigma^2(M)$ is the variance of the density field associated with mass scale $M$. Halo formation is mapped onto the first-crossing of a collapse barrier $B(S,z)$ by these trajectories. The first-crossing distribution gives the probability that a region of a given mass has collapsed by a specified time, and therefore determines the abundance of haloes of that mass. By conditioning on the known descendant halo, the first-crossing distribution can be used to determine the probability distribution of progenitor halo masses at earlier times. These first-crossing distributions therefore provide the basis for constructing Monte Carlo realisations of merger tree histories without the computational cost of explicitly following halo populations in cosmological $N$-body simulations.

A variety of algorithms have been developed to construct merger trees from EPS progenitor distributions \citep[e.g.][]{LaceyCole1993,ColeEtAl2000,Zentner2007}. Among the most widely used is the prescription of \citet{ParkinsonColeHelly2008}, hereafter PCH08. PCH08 modifies the EPS progenitor distribution with an empirical multiplicative correction calibrated against merger trees extracted from the Millennium simulation \citep[][]{Springel_2005}. This substantially improves agreement with simulated halo assembly statistics while retaining the speed and flexibility of a Monte Carlo approach. However, constructing a merger tree requires more than specifying the progenitor distribution: it must be sampled to generate individual progenitor haloes and their masses, which are then linked across successive timesteps to construct the history of mergers and accretion. Different choices in these steps can produce different merger tree statistics even when they are based on the same underlying excursion set model \citep{ZhangFakhouriMa2008,JiangVanDenBosch2014}.

This distinction between the physical excursion set model and the numerical algorithm used to construct merger trees becomes particularly important when moving beyond the standard spherical collapse treatment \citep[][]{GunnGott1972}. In the excursion set picture, the collapse barrier specifies the density contrast that a region must reach in order to collapse into a halo at a given redshift. For spherical collapse, this threshold depends on redshift but not on the smoothing scale, giving a barrier of the form $B(S,z)\equiv B(z)=\delta_{\rm c}(z)$. As the smoothing scale is varied, the density contrast trajectory $\delta(S)$ therefore crosses a fixed threshold at each redshift.  For the commonly used sharp-$k$ filter, which includes Fourier modes up to a specified wavenumber and excludes modes at larger wavenumbers, i.e. smaller spatial scales, changing the smoothing scale adds independent sets of Fourier modes to the density field. The resulting trajectories are therefore Markovian random walks: the statistics of the next step depend only on the current value of the trajectory. For the constant spherical-collapse barrier, this allows the first-crossing distribution to be obtained analytically.

More general models allow the collapse threshold to be a moving barrier that depends on smoothing scale $S$. Physically, this means that the density contrast required for collapse depends on the mass scale of the region being considered. Ellipsoidal collapse provides a familiar example \citep{ShethTormen2002}. Merger tree calculations incorporating moving barriers have been explored previously \citep{HiotelisDelPopolo2006}. The challenge is then to determine the first-crossing distribution for the chosen barrier and incorporate it consistently into the merger tree construction.

For a Markovian walk (\S\ref{sec:variance_barrier}), \citet{ZhangHui2006} derived an integral equation for the first-crossing distribution of a general scale-dependent barrier. Their formulation follows from probability conservation and allows the first-crossing problem to be solved numerically without explicitly generating ensembles of random walks. \citet{BensonEtAl2013} subsequently developed a numerically robust implementation and applied the approach to generalised EPS merger histories beyond the canonical Cold Dark Matter (CDM) model. More recently, \citet{NadlerEtAl2023} demonstrated how related first-crossing calculations can be incorporated into practical merger tree algorithms, including merger histories conditioned to satisfy a prescribed progenitor constraint. The distinction is therefore between calculating the statistical distribution of first crossings and sampling that distribution to construct individual merger histories.

This separation is particularly useful for non-standard dark matter models, in which halo formation can differ from the CDM case through two distinct effects. First, changes to the linear matter power spectrum, $P(k)$, alter the abundance of fluctuations on different mass scales, which propagate directly into the relation between halo mass $M$ and excursion set variance $S(M)$. Second, the physics governing collapse may change; if collapse is suppressed or delayed in a scale-dependent manner, the critical overdensity and hence the barrier $B(S,z)$ can also change. Altering the linear power spectrum alone does not describe such a modification of the collapse dynamics; a consistent excursion set treatment must therefore allow $P(k)$ and $B(S,z)$ to be modified independently.

A flexible merger tree framework should therefore keep several ingredients explicit: the linear density field characterised by \(P(k)\), which determines \(S(M)=\sigma^2(M)\); the smoothing prescription and its mapping between smoothing scale and mass; the collapse model, which determines \(B(S,z)\); and the first-crossing calculation, which determines the progenitor distribution. Separating these ingredients allows changes in cosmology and collapse physics to be propagated through the same merger tree machinery, without requiring a new tree-building algorithm for each model. It also provides a common interface between the calculation of progenitor statistics and the construction of merger trees, allowing the latter to remain largely agnostic to the underlying cosmology and collapse physics.

A general first-crossing solver of this kind already exists within \texttt{Galacticus} \citep{Benson2012}, which implements the formulation of \citet{ZhangHui2006} as part of a comprehensive semi-analytic galaxy formation framework. \texttt{Galacticus} provides a broad modelling environment, whereas a standalone implementation can instead prioritise fast generation of Monte Carlo halo merger trees and straightforward integration with existing analysis workflows.

This motivates \texttt{foraois}, an open-source Python framework designed to provide fast and flexible generation of halo merger histories across different cosmologies and collapse models. The framework implements both the PCH08 prescription and the general Markovian first-crossing solver of \citet{ZhangHui2006}, together with constrained excursion histories based on the Brownian-bridge construction of \citet{NadlerEtAl2023}.  The same machinery can consequently be used to generate merger trees for different cosmologies, dark matter models, and, in the case of the Zhang--Hui formulation, for different collapse prescriptions (for merger trees, through the serial builder only; \S\ref{sec:architecture}). Its fast generation of large ensembles of merger trees makes it suitable for applications in which halo histories must be repeatedly generated. This includes coupling to semi-analytical models of galaxy formation; we illustrate this  using the \texttt{Ashvini} model \citep{MenonPower2024,MenonBaluPower2026}.

The remainder of this paper is structured as follows. In \S~\ref{sec:framework}, we describe the mathematical framework underlying \texttt{foraois}. In \S~\ref{sec:implementation}, we describe its software implementation, including the treatment of different dark matter cosmologies and the decomposition of halo growth into resolved mergers and smooth accretion. We test the first-crossing solver and tree builders in \S~\ref{sec:validation} and characterise their computational performance in \S~\ref{sec:performance}. In \S~\ref{sec:ashvini}, we use the \texttt{Ashvini} semi-analytic model to compute a stellar-to-halo mass relation from \texttt{foraois} trees for three dark matter models. We discuss the interpretation and limitations of the framework in \S~\ref{sec:discussion} and summarise our conclusions in \S~\ref{sec:summary}.

\section{Mathematical framework}
\label{sec:framework}

The following subsections provide a brief summary of the elements of the extended Press--Schechter (EPS) formalism \citep{PressSchechter1974,BondEtAl1991,LaceyCole1993} needed to describe the merger tree algorithms implemented in \texttt{foraois}. For a more comprehensive review, we refer the interested reader to \citet{Zentner2007} and references therein.

\subsection{Variance and collapse barrier}
\label{sec:variance_barrier}

For a given linear matter power spectrum $P(k)$, the variance of the density field smoothed on a characteristic filter scale $R$ is
\begin{equation}
S(R)\equiv\sigma^2(R)=\frac{1}{2\pi^2}\int_0^\infty k^2P(k)W^2(kR)\mathrm{d}k ,
\label{eq:variance}
\end{equation}
where $W(kR)$ is the Fourier transform of the smoothing window. Relating the variance to halo mass additionally requires a mapping between filter scale and mass. For a real space top-hat filter,
\begin{equation}
M=\frac{4\pi}{3}\bar{\rho}_{\rm m}R^3 ,
\label{eq:mass_radius}
\end{equation}
where $\bar{\rho}_{\rm m}$ is the mean comoving matter density. The choices adopted in \texttt{foraois} are described in \S~\ref{sec:dm_models}.

In excursion set theory, the smoothed linear density contrast traces a trajectory $\delta(S)$ as the smoothing scale is varied. Halo formation is associated with the first crossing of a collapse barrier by the density contrast trajectory, $\delta=B(S,z)$, which specifies the linearly extrapolated overdensity required for collapse at variance $S$ and redshift $z$. For spherical collapse, the barrier is independent of $S$,
\begin{equation}
B(S,z)=\delta_{\rm c}(z)\simeq\frac{\delta_{\rm sc}}{D(z)},
\label{eq:spherical_barrier}
\end{equation}
where $\delta_{\rm sc}\simeq1.686$ and $D(z)$ is the linear growth factor. More general collapse prescriptions can produce a scale-dependent, or \emph{moving}, barrier, $B(S,z)\neq B(z)$.

The general first-crossing solver of \citet{ZhangHui2006} assumes a Markovian excursion-set walk. This is realised by a sharp-$k$ filter, for which increasing $S$ adds statistically independent Fourier modes \citep{BondEtAl1991}, so the walk has independent increments. The variance $S(M)=\sigma^2(M)$ is set by the linear power spectrum, the window function, and the mapping between smoothing scale and halo mass. Keeping this relation explicit lets the same first-crossing calculation be applied to different cosmologies and smoothing prescriptions. The solver is exact for a Markov (Brownian) walk. The default \texttt{foraois} configurations instead use a real-space top-hat window, for which the walks are correlated; applying the Markov solver to the corresponding $S(M)$ is then an approximation, and no non-Markovian corrections are implemented \citep[see e.g.][]{MaggioreRiotto2010}.

\subsection{Conditional first-crossing distributions}
\label{sec:conditional_first_crossing}

Consider a descendant halo of mass $M_0$ at redshift $z_0$, with
\begin{equation}
S_0=S(M_0),
\qquad
B_0=B(S_0,z_0).
\end{equation}
At an earlier redshift $z_1>z_0$, a progenitor of mass $M_1<M_0$ corresponds to a larger variance, $S_1>S_0$. Shifting the origin of the excursion-set trajectory from the descendant point $(S_0,B(S_0,z_0))$ to the origin, and defining $s\equiv S-S_0$, gives the conditional barrier
\begin{equation}
b(s;z_1\mid S_0,z_0)
\equiv
B(S_0+s,z_1)-B(S_0,z_0).
\label{eq:shifted_barrier}
\end{equation}
The conditional first-crossing problem is therefore that of a random walk starting at the origin and crossing $b(s)$. If $f(s)$ denotes the corresponding first-crossing distribution, then $f(s)\,\mathrm{d}s$ is the fraction of the descendant mass $M_0$ assigned to progenitors with variance in $\mathrm{d}S_1=\mathrm{d}s$ about $S_1=S_0+s$. The mass in these progenitors is therefore $M_0 f(S_1-S_0)\,\mathrm{d}S_1$. 

Dividing by the progenitor mass $M_1$ converts this mass fraction into an expected number of progenitors,
\begin{equation}
\mathrm{d}N=\frac{M_0}{M_1}f(S_1-S_0)\mathrm{d}S_1.
\end{equation}
Finally, changing variables from $S_1$ to $M_1$ gives the conditional progenitor mass function
\begin{equation}
\frac{\mathrm{d}N}{\mathrm{d}M_1}
=\frac{M_0}{M_1}f(S_1-S_0)
\left|
\frac{\mathrm{d}S_1}{\mathrm{d}M_1}
\right|.
\label{eq:conditional_mass_function}
\end{equation}

\subsection{Constant and moving barriers}
\label{sec:eps_solution}

For a spherical-collapse barrier, the effective conditional barrier is constant,
\begin{equation}
b(s)=\Delta B=\delta_{\rm c}(z_1)-\delta_{\rm c}(z_0),
\end{equation}
and the conditional first-crossing distribution has the standard analytic EPS form \citep{BondEtAl1991,LaceyCole1993},
\begin{equation}
f_{\rm EPS}(s,\Delta B)=
\frac{\Delta B}
{\sqrt{2\pi},s^{3/2}}
\exp\left[
-\frac{(\Delta B)^2}{2s}
\right].
\label{eq:eps_conditional}
\end{equation}
For a general moving barrier, the effective barrier $b(s)$ is not constant and the first-crossing distribution does not generally have a closed analytic form. \texttt{foraois} therefore implements the probability-conservation formulation of \citet{ZhangHui2006}, which determines the first-crossing distribution for a specified barrier and a Markovian walk.

The resulting first-crossing distribution provides one route from the excursion-set model to a merger-tree algorithm. A different route is provided by the PCH08 prescription \citep{ParkinsonColeHelly2008}, which modifies the EPS progenitor-splitting rate using a fitting function calibrated against $N$-body merger trees. This means that both approaches generate stochastic progenitor histories but do not represent the same approximation at the same stage of the calculation. The Zhang--Hui approach determines the first-crossing distribution directly from the specified barrier, whereas PCH08 modifies the EPS tree-splitting prescription empirically.

This distinction is important when comparing merger trees generated by the two methods. For Zhang--Hui, changing the collapse barrier changes the first-crossing distribution and therefore the progenitor statistics supplied to the tree sampler. For PCH08, the calibrated splitting prescription is applied to the EPS progenitor statistics and is not, in the present implementation, a general solver for an arbitrary moving barrier. The two methods therefore share the same cosmological mapping from $M$ to $S(M)$ and the same overall tree representation, but differ in how progenitor statistics are obtained.

The framework separates the linear matter power spectrum from the collapse prescription. Changes to the linear matter power spectrum modify $S(M)$, while changes to the collapse prescription modify $B(S,z)$. These inputs can therefore be varied independently in the Zhang--Hui first-crossing calculation (and, for merger trees, its serial builder; \S\ref{sec:architecture}), while the PCH08 prescription retains its calibrated modification of the EPS splitting statistics.

\subsection{From progenitor statistics to merger trees}
\label{sec:tree_sampling}

The excursion-set calculation provides a statistical description of progenitor growth, while a merger tree is a particular stochastic realisation of that description. In both cases the continuous statistical description must be converted into a finite set of resolved progenitors at each timestep. The two prescriptions implemented in \texttt{foraois} obtain the progenitor statistics differently and consequently use different sampling probabilities.

The Zhang--Hui algorithm samples binary mass splits from the conditional first-crossing distribution $f(S_1-S_0)$ of the specified barrier, evaluated in the Markov approximation. For a descendant of mass $M_0$ at redshift $z_0$, with $S_i=\sigma^2(M_i)$, the corresponding progenitor mass function at $z_1$ is
\begin{equation}
\frac{\mathrm{d}N}{\mathrm{d}M_1}(M_1,z_1\mid M_0,z_0)
=
\frac{M_0}{M_1}f(S_1-S_0)
\left|
\frac{\mathrm{d}S_1}{\mathrm{d}M_1}
\right| .
\end{equation}
In practice a progenitor is drawn directly from the cumulative distribution of $f$ in $S$, restricted to $M_1\in[M_{\rm res},\,M_0(1-F)-M_{\rm res}]$ so that both fragments of a split are resolved \citep[following][]{NadlerEtAl2023}. The probability of a resolved split in a step is the integral of $f$ over this range.

PCH08 instead modifies the EPS progenitor-splitting rate with an empirically calibrated factor,
\begin{equation}
\frac{\mathrm{d}N}{\mathrm{d}M_1}
=
G(\sigma_1/\sigma_0,\,\delta_c/\sigma_0)\,\frac{\mathrm{d}N_{\rm EPS}}{\mathrm{d}M_1},
\end{equation}
with
\begin{equation}
G=G_0\left(\frac{\sigma_1}{\sigma_0}\right)^{\gamma_1}\left(\frac{\delta_c}{\sigma_0}\right)^{\gamma_2};
\end{equation}
here $\sigma_i=\sigma(M_i)$, $\delta_c=\delta_c(z_0)$, and $(G_0,\gamma_1,\gamma_2)=(0.57,0.38,-0.01)$ are calibrated against $N$-body merger trees \citep{ParkinsonColeHelly2008}. The timestep enters through the barrier shift $\Delta\omega=\delta_c(z_1)-\delta_c(z_0)$ multiplying the rate. Splits are drawn by rejection sampling against an upper-bound rate $N_{\rm upper}$, and no split is possible when $M_0<2M_{\rm res}$.

In both cases, progenitors with mass below a minimum resolved mass $M_{\rm res}$ are not individually represented; their contribution is treated as smooth accretion, a fraction $F$ of the descendant mass. For Zhang--Hui, $F$ is the tail of $f$ beyond $S_{\rm res}=\sigma^2(M_{\rm res})-S_0$; for PCH08 it is the integral of the modified rate over unresolved progenitors. The mass is then conserved exactly by construction,
\begin{equation}
M_0=\sum_i M_{1,i}+F\,M_0 .
\end{equation}
Both algorithms allow at most one split per timestep, so the step must be small enough that the expected number of splits is $\lesssim 0.1$. Resolved progenitors are propagated to the preceding timestep and the procedure is repeated recursively.

\subsection{Time stepping and progenitor sampling}
\label{sec:tree_timestep}

The Zhang--Hui tree generator converts the first-crossing distribution into a binary sequence of resolved progenitor splits. For a descendant of mass $M_0$, the cumulative first-crossing distribution is evaluated over the resolved mass range, while crossings corresponding to progenitors below $M_{\rm res}$ are represented as smooth accretion. The unresolved fraction is
\begin{equation}
F_{\rm unres}
=
C(s_{\max})-C(s_{\rm res}),
\end{equation}
with the result restricted to the physical range $0\leq F_{\rm unres}\leq1$. Here
\begin{equation}
C(s)=\int_0^s f(u),\mathrm{d}u,
\end{equation}
$s_{\rm res}=S(M_{\rm res})-S(M_0)$, and $s_{\max}=8s_{\rm res}$ by default. The mass available for resolved progenitors is then
\begin{equation}
M_{\rm avail}=M_0(1-F_{\rm unres}).
\end{equation}

If $M_{\rm avail}>2M_{\rm res}$, the binary sampler draws the mass of one progenitor from the restricted first-crossing distribution and assigns the remaining resolved mass to the other. No ordering of the two is imposed: the drawn mass can exceed the complementary mass. If $M_{\rm avail}\leq2M_{\rm res}$, no resolved split is made and the resolved mass is propagated as a single progenitor. The drawn progenitor mass lies in
$$
M_1\in[M_{\rm res},M_{\rm avail}-M_{\rm res}],
$$
so that both retained progenitors satisfy the adopted mass-resolution limit.

This procedure is an approximation to the continuum progenitor mass function in equation~\ref{eq:conditional_mass_function}: rather than independently sampling a number of progenitors from $\mathrm{d}N/\mathrm{d}M_1$, it samples a binary mass split while assigning the unresolved component to smooth accretion. The resulting progenitors are then used as descendants for the preceding timestep, and the procedure is repeated recursively.

Both tree builders allow at most one resolved split per timestep, so the timestep must be sufficiently small that multiple resolved splits are unlikely. For PCH08 we use its adaptive envelope criterion $N_{\rm upper}\lesssim0.1$; for the Zhang--Hui binary sampler we use the corresponding EPS expected split count $E\lesssim0.1$ as a practical criterion. These criteria control the timestep approximation rather than providing convergence or physical validation. Note that a binary split is not the only choice; for example, \citet{KauffmannWhite1993} draw several progenitors per timestep from the conditional mass function. \citet{NeisteinDekel2008} argue that a binary-merger interpretation of the EPS conditional mass function is inconsistent because a typical merger involves more than two progenitors, even in the small-timestep limit; this contrasts with N-body simulations, where mergers resolve into binary events at short enough timesteps. \citet{NeisteinDekel2008} derived revised EPS merger rates that account for multiple progenitors, matched to N-body merger rates, and they used the resulting rates in a Monte Carlo tree algorithm that recovers the EPS progenitor mass function. We do not implement multiple-progenitor sampling here, and we use the binary sampler only together with the timestep criteria above, under which multiple resolved splits per step are unlikely. The effects of timestep, mass resolution and numerical first-crossing resolution are examined separately in Section~\ref{sec:validation}.


\section{Software implementation}
\label{sec:implementation}

\subsection{Architecture}
\label{sec:architecture}

The separation described in \S~\ref{sec:framework} is reflected directly in the structure of \texttt{foraois}. The implementation has three principal layers: (i) cosmological calculations that determine the variance--mass relation $S(M)$ from a linear matter power spectrum, (ii) collapse-barrier models that supply the collapse threshold as a function of mass and redshift, and (iii) first-crossing and tree-building algorithms that operate on these derived quantities. Cosmology-dependent quantities, including $P(k)$, $S(M)$, the growth factor $D(z)$ and the collapse threshold, are calculated at initialisation and stored in reusable lookup structures. The tree-generation algorithms therefore operate on common variance and barrier interfaces rather than directly on the underlying cosmological model. Changing the power spectrum or smoothing prescription changes the derived inputs without requiring changes to the tree-construction algorithms, while alternative collapse prescriptions can be supplied through the barrier interface.

The barrier interface accepts mass- and redshift-dependent collapse thresholds, and the numerical first-crossing solver and serial Zhang--Hui builder accept an arbitrary supplied barrier for a Markovian walk without modification. The vectorised, compiled and constrained tree-building paths currently require a flat barrier, and PCH08 is likewise restricted to the flat spherical-collapse barrier; a scale-dependent barrier is therefore available for merger trees through the serial Zhang--Hui builder only. The release includes an illustrative scale-dependent barrier, linear in the variance, for testing this path. No moving-barrier merger-tree calculation is demonstrated in this paper: moving barriers are tested only at the level of the first-crossing solver (\S\ref{sec:validation}). The present release provides mass-independent spherical-collapse barriers for CDM and WDM and a mass-independent placeholder barrier for FDM.

Three tree-generation strategies are provided:

\begin{enumerate}
\item The \textbf{PCH08 backend} implements the branching-rate and rejection-sampling construction of \citet{ParkinsonColeHelly2008}.

\item The \textbf{Zhang--Hui backend} implements the first-crossing formalism of \citet{ZhangHui2006}. The serial implementation evaluates the first-crossing solution numerically for a supplied barrier and a Markovian walk, whereas the vectorised and compiled implementations use the corresponding analytic solution for the currently supported flat-barrier models, enabling efficient generation of large ensembles.

\item The \textbf{constrained backend} extends the first-crossing calculation using a Brownian bridge to generate conditioned main-branch histories \citep{NadlerEtAl2023}.
\end{enumerate}

The PCH08 backend constructs full branching trees containing resolved progenitor nodes. The Zhang--Hui and constrained backends provide main-branch histories, with ensemble representations retaining the resolved merger and smooth-accretion channels. The population-comparison tools can recursively evolve Zhang--Hui progenitors when full progenitor populations are required. These output representations are consequently explicit in the API rather than inferred from a common tree object.

The scientific algorithms are separated from their computational implementations. PCH08 main-branch generation is available in serial, vectorised and compiled thread-parallel forms. Zhang--Hui provides a general serial numerical implementation together with vectorised and compiled constant-barrier implementations. The constrained sampler currently operates serially. The Python API and command-line interface expose these methods while selecting the appropriate computational implementation.

The principal modules and their responsibilities are summarised in Table~\ref{tab:modules}.

\begin{table*}[t]
\caption{Principal \texttt{foraois} modules and their responsibilities.}
\label{tab:modules}
\begin{tabular}{p{0.30\textwidth}p{0.65\textwidth}}
\hline
Module & Responsibility \\
\hline
\texttt{cosmo\_utils} &
Linear growth factor, collapse threshold, cosmic time, $P(k)$ and $S(M)$; constructs cosmology-dependent derived quantities \\

\texttt{transfer\_functions} &
CDM, WDM and FDM transfer functions (\S~\ref{sec:dm_models}) \\

\texttt{collapse} &
Collapse barrier $\texttt{delta\_c}(M,z,\texttt{cosmo\_data})$, selected by the \texttt{barrier} configuration (\texttt{fixed} or \texttt{linear}); the general first-crossing solver and serial Zhang--Hui builder accept a scale-dependent barrier \\

\texttt{pch\_trees} &
PCH08 branching-rate and rejection-sampling tree construction \\

\texttt{first\_crossing} &
General first-crossing solver for a specified barrier and a Markovian walk, including the constant-barrier limit following \citet{ZhangHui2006} \\

\texttt{zhang\_hui\_trees} &
Tree construction based on Zhang--Hui first-crossing distributions, with serial numerical and vectorised/compiled constant-barrier implementations \\

\texttt{first\_crossing\_constrained} &
Constrained first-crossing and Brownian-bridge calculations following \citet{NadlerEtAl2023} \\

\texttt{zhang\_hui\_constrained\_trees} &
Constrained tree construction using the Brownian-bridge machinery \\

\texttt{tree\_algorithm} &
Structural interface shared by the PCH08 and unconstrained general-first-crossing backends \\

\texttt{diagnostics} &
Step-size and sampling-consistency diagnostics; mass conservation is checked in the tree test suite \\

\texttt{utils.plot} &
Tree, mass-function and merger-history visualisation \\
\hline
\end{tabular}
\end{table*}

The repository includes automated tests and executable validation scripts for the cosmological calculations, first-crossing solvers, merger-tree construction and constrained histories (306 tests at the time of writing). The numerical validation of these components is described in \S~\ref{sec:validation}.

\subsection{Basic package interaction}
\label{sec:usage}

The standard workflow is to load a YAML configuration, construct a \texttt{CosmoData} object and initialise a tree generator. The following example builds one main-progenitor history for a Planck-parameter CAMB cosmology.

\begin{lstlisting}
from foraois import CosmoData, PCHMergerTree
from foraois.utils import io

run_params = io.get_params(
"config/planck2018_camb.yml"
)
cosmo_data = CosmoData(
run_params,
redshift=[0.0]
)
tree_generator = PCHMergerTree(
cosmo_data,
run_params
)

tree = tree_generator.build_tree(
M0=1.0e12,
z0=0.0,
z_max=2.0,
M_res=1.0e10,
dz=0.2,
)
\end{lstlisting}

The same structural tree-building interface is used by the other tree-generation methods. \texttt{CosmoData}, \texttt{PCHMergerTree}, \texttt{ZhangHuiMergerTree}, \texttt{build\_constrained\_tree} and \texttt{TreeAlgorithm} are exported from the top-level \texttt{foraois} namespace. A \texttt{ZhangHuiMergerTree} instance is initialised with its backend-specific numerical options before using the tree-building interface. The constrained entry point additionally specifies the target $(M_1,z_1)$ used to condition the history. The dark-matter model is selected through the cosmological configuration; changing to a WDM or FDM configuration therefore changes the derived cosmological inputs without requiring a different tree representation.

\begin{figure}[t]
\centering
\includegraphics[width=\linewidth]{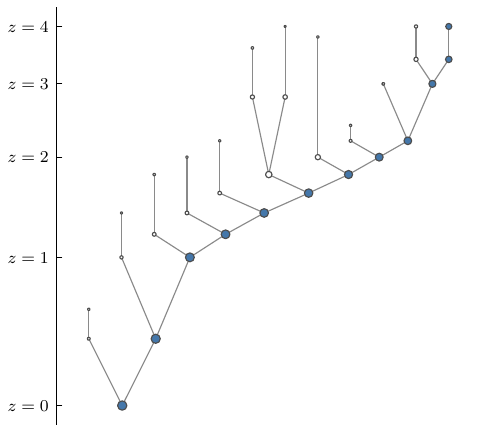}
\caption{A full branching Monte Carlo merger tree rendered with the dendrogram utility. This representation is produced by \texttt{PCHMergerTree.build\_full\_tree}, as used in \texttt{scripts/make\_tree\_dendrogram\_figure.py}, rather than by the main-branch method in \S~\ref{sec:usage}. Branch points mark resolved progenitor splits, node area scales with the logarithm of the halo mass, and the main branch is filled. The vertical axis is the logarithm of cosmic time, labelled by redshift. We assume a halo mass of $M=10^{13}\,M_{\odot}/h$, $M_{\rm res}=10^{11}\,M_\odot/h$, and $dz=0.2$.}
\label{fig:tree_dendrogram}
\end{figure}

Configuration can also be supplied directly rather than through YAML. \texttt{CosmoData.from\_params(...)} accepts cosmological parameters as keyword arguments, as described in \S~\ref{sec:usage_extras}. Once a tree has been generated, \texttt{utils.plot} provides utilities for visualising its structure, including the dendrogram shown in Figure~\ref{fig:tree_dendrogram}.

The repository also contains executable validation and figure-generation scripts, including \texttt{scripts/make\_validation\_figure.py} and \texttt{scripts/validate\_zhang\_hui\_vs\_pch08.py}, together with an introductory notebook. These provide examples of the analysis workflow beyond the minimal tree-generation example. Reproduction of a particular manuscript figure additionally requires the source revision, run settings and output data; the supplied manuscript materials are discussed in \S~\ref{sec:availability}.

\subsection{Configuration without YAML and optional dependencies}
\label{sec:usage_extras}

YAML configuration is convenient for reproducible file-based runs, but is not required. \texttt{CosmoData.from\_params(...)} constructs a \texttt{CosmoData} instance directly from cosmological parameters supplied as keyword arguments, including \texttt{H0}, \texttt{OmegaM}, \texttt{OmegaBar}, \texttt{As}, \texttt{ns} and \texttt{tau}. Direct construction is particularly useful when cosmological or dark-matter parameters are generated programmatically, for example for parameter grids or inference calculations.

The core installation requires only \texttt{numpy}, \texttt{scipy}, \texttt{pyyaml} and \texttt{matplotlib}. \texttt{numba} and the CLASS/CAMB power-spectrum interfaces are optional dependencies, required only for the corresponding compiled or power-spectrum functionality.

\subsection{Cosmological models and dark matter}
\label{sec:dm_models}

Because the tree-building kernels use the derived variance $S(M)$ and collapse barrier $B(S,z)$, dark-matter physics enters the framework through the ingredients used to construct these quantities. In general, a dark-matter model may modify the linear power spectrum, the smoothing prescription, the collapse barrier, or some combination of these. The present release provides CDM, thermal-relic WDM and FDM linear transfer functions, together with optional sharp-$k$ smoothing for WDM and FDM.

\paragraph{Linear power spectra}

\begin{figure*}[t]
\centering
\includegraphics[width=\linewidth]{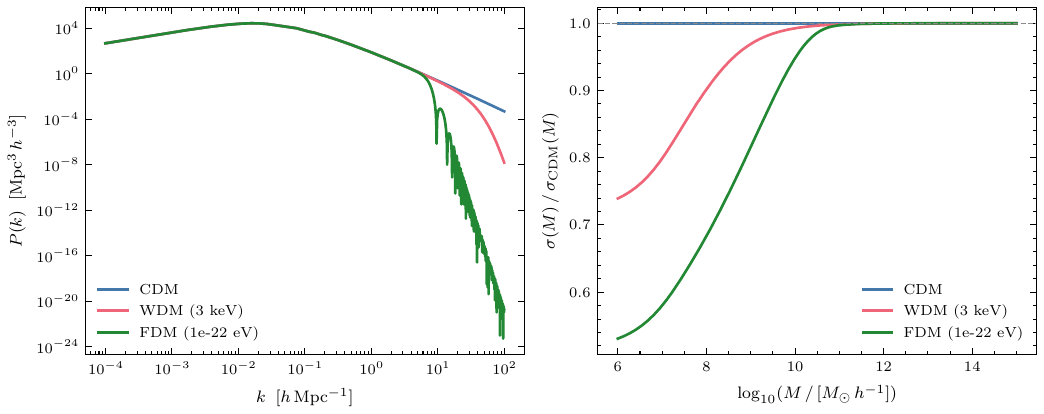}
\caption{Linear power spectra (left) and rms fluctuation ratios (right) $\sigma(M)/\sigma_{\rm CDM}(M)$ for CDM, a $3\,{\rm keV}$ thermal-relic WDM model and $10^{-22}\,{\rm eV}$ FDM, evaluated through \texttt{CosmoData} on a Planck 2018 CAMB cosmology. The default configurations used by the figure script differ in their transfer functions and use the same real-space top-hat filter.}
\label{fig:dm_model_comparison}
\end{figure*}

For all models, the dark-matter power spectrum is constructed from the CDM spectrum as
\begin{equation}
P_{\rm DM}(k)
=
T^2(k)P_{\rm CDM}(k).
\end{equation}
For CDM,
\begin{equation}
T_{\rm CDM}(k)=1.
\end{equation}

For thermal-relic WDM, we use the fitting form of \citet{VielEtAl2005},
\begin{equation}
T_{\rm WDM}(k)
=\left[1+(\alpha_{\rm WDM}k)^{2\nu}\right]^{-5/\nu},
\label{eq:wdm_transfer}
\end{equation}
with $\nu=1.12$ and
\begin{equation}
\alpha_{\rm WDM}
=0.049
\left(\frac{m_{\rm WDM}}{\mathrm{keV}}\right)^{-1.11}
\left(\frac{\Omega_{\rm m}}{0.25}\right)^{0.11}
\left(\frac{h}{0.7}\right)^{1.22}
h^{-1},\mathrm{Mpc}.
\label{eq:wdm_alpha}
\end{equation}
The fitting form describes thermal-relic free-streaming suppression. In the present implementation, its density argument is the total matter density parameter $\Omega_{\rm m}$, including baryons, rather than a separately supplied $\Omega_{\rm WDM}$.

We note that this calibration was derived for particle masses below approximately 1.5 keV. \citet{VogelAbazajian2023} re-fitted the same functional form at higher thermal-relic masses and find less suppression at fixed particle mass than the \citet{VielEtAl2005} form. For our adopted cosmology, the two fits agree near 1 keV, while the half-mode wavenumber (where the transfer function falls to one half) is higher in the \citet{VogelAbazajian2023} fit by approximately 10 per cent at 3 keV and approximately 20 per cent at 10 keV. Beyond the cutoff, the \citet{VogelAbazajian2023} transfer function is higher by a factor of approximately 2–4 at three times the half-mode wavenumber for 3–10 keV. These values are approximate and are obtained directly from the two published fitting formulae, not from a Boltzmann-code calculation. Because the functional form is the same, the difference enters only through the values of $\alpha_{\rm WDM}$ and $\nu$. This alternative calibration is not used for the results in this paper, but it will be added to foraois as an alternative option. The WDM results shown here should therefore be read as illustrative of the framework rather than as a precise calibration for a given particle mass.

For FDM, we use the transfer-function approximation of \citet{HuBarkanaGruzinov2000},
\begin{equation}
T_{\rm FDM}(k)
=
\frac{\cos(x^3)}{1+x^8},
\label{eq:fdm_transfer}
\end{equation}
where
\begin{equation}
x
=
1.61
\left(
\frac{m_a}{10^{-22},\mathrm{eV}}
\right)^{1/18}
\frac{k}{k_{J,\rm eq}}
\end{equation}
and
\begin{equation}
k_{J,\rm eq}
\simeq
9
\left(
\frac{m_a}{10^{-22},\mathrm{eV}}
\right)^{1/2}
\mathrm{Mpc}^{-1}.
\end{equation}

\paragraph{Smoothing}

For WDM and FDM, a sharp-$k$ window is available in place of the conventional real-space top hat. This suppresses the contribution of modes above a sharp Fourier-space cutoff rather than retaining the extended Fourier-space tails of the real-space top-hat filter. The sharp-$k$ filter requires an explicit mass--filter assignment; the implementation uses
\begin{equation}
k_{\rm c}
=
\frac{c_{\rm sk}}{R},
\qquad
c_{\rm sk}=2.5,
\end{equation}
following the sharp-$k$ calibration used in related WDM and FDM halo-mass function calculations \citep{BensonEtAl2013,KulkarniOstriker2022}. Because the window is discontinuous, $S(M)$ and its derivative are evaluated using expressions appropriate to the sharp-$k$ case. The results shown in this paper use the real-space top-hat window for all three models. For WDM and FDM this retains power above the cutoff through the filter's Fourier-space tails, so the sharp-$k$ window is the more appropriate choice for quantitative studies of those models. The sharp-$k$ filter is also the construction for which the walk is Markovian (\S\ref{sec:variance_barrier}).

Figure \ref{fig:dm_model_comparison} shows the linear power spectrum $P(k)$ (left) for CDM, WDM (corresponding to a 3 keV thermal-relic), and FDM ($10^{-22}$ eV). The panel on the right shows the ratio of the rms fluctuation $\sigma(M)$ to the case for CDM. The figures were generated with the CAMB backend of \foraois.

\paragraph{Collapse barriers}

The transfer function and smoothing prescription determine the linear variance entering the excursion-set calculation, but they do not uniquely determine the collapse barrier. The present implementation uses a mass-independent WDM barrier and a mass-independent placeholder for FDM. The default WDM configuration therefore represents free-streaming suppression through the linear power spectrum without including a separately calibrated scale-dependent collapse threshold.

This distinction is relevant for WDM because residual velocity dispersion can also affect collapse. \citet{BensonEtAl2013}, for example, considered both the WDM power-spectrum cutoff and a scale-dependent effective collapse barrier. The general barrier interface in \texttt{foraois} allows such alternatives to be investigated independently of the transfer function in the general first-crossing solver and the serial Zhang--Hui builder, but they are not part of the present default WDM model. Likewise, the current FDM implementation does not provide an independently calibrated moving-collapse barrier.

The resulting architecture treats the dark-matter model as a modular combination of linear power spectrum, smoothing prescription and collapse model.

\subsection{Accretion decomposition: smooth accretion and mergers}
\label{sec:accretion_decomposition}

\begin{figure}[t]
\centering
\includegraphics[width=\linewidth]{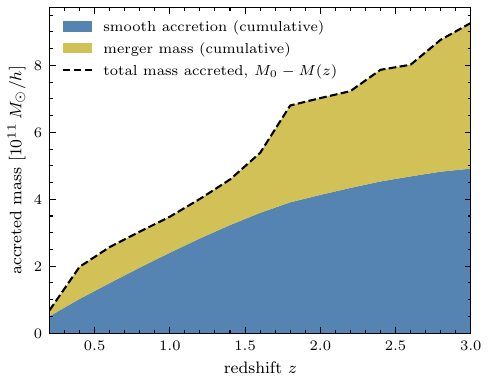}
\caption{Cumulative smooth-accretion and resolved-merger contributions to the main-branch mass history of a representative PCH08 tree, decomposed on retained resolved steps through equation~\ref{eq:smooth_accretion}. The two channels together reproduce the corresponding main-branch growth while retaining information that cannot be recovered uniquely from a differenced mass history alone.}
\label{fig:smooth_vs_merger}
\end{figure}

On a resolved main-branch step, the parent mass can be divided into the continuing progenitor, resolved secondary progenitors and an unresolved component. The ensemble interfaces expose separate smooth-accretion and merger channels. These fields are also available in serial Zhang--Hui and constrained histories; for serial PCH08 output they can instead be reconstructed from the returned parent and progenitor masses.

We define the merger contribution as the summed mass of resolved progenitors other than the most massive continuing branch,
\begin{equation}
\Delta M_{\rm merger}=\sum_{i\neq{\rm main}}M_i,
\end{equation}
and the smooth component as the conservation residual
\begin{equation}
\Delta M_{\rm smooth}
=M_{\rm parent}-\max_i(M_i)-\Delta M_{\rm merger}.
\label{eq:smooth_accretion}
\end{equation}
Equation~\ref{eq:smooth_accretion} expresses mass conservation on retained resolved steps. When the continuing branch falls below $M_{\rm res}$, its mass and the corresponding accretion channels are no longer retained, so the identity does not apply to the terminal transition below the resolution limit. Figure \ref{fig:smooth_vs_merger} shows the decomposition into smooth-accretion and mergers for the cumulative accretion history for a  $10^{11}h^{-1}\Msun$ halo. The particular example has been generated with the PCH08 backend of \foraois.

The decomposition is exposed at the point where progenitor masses are generated, allowing downstream models to apply separate prescriptions to smooth growth and merger-driven growth without reconstructing the distinction from a differenced mass history.

For the constrained branch described by \citet{NadlerEtAl2023}, the same fields are retained. Discrete secondary subtrees are not generated along the constrained portion in the current implementation, so the merger field is zero there; smooth growth records the mass change between successive running-maximum points of the sampled constrained excursion. This limitation is discussed in \S~\ref{sec:discussion}.


\section{Validation}
\label{sec:validation}

The validation suite separates analytic first-crossing limits, direct Monte Carlo checks, checks of the PCH08 rate machinery, tree-level comparisons, and tests of the constrained extension. These address different questions: accuracy of the first-crossing density does not establish that a binary sampler reproduces the continuum progenitor-number distribution, and agreement between a sampler and a quadrature of its own rate is an internal check.

\subsection{Closed-form limits}

The general first-crossing solver of \citet{ZhangHui2006} is tested on two cases for which analytic solutions are available - a flat barrier and a barrier linear in variance. For the tested grids, the flat barrier solution agrees with the analytic EPS result to floating-point precision, while the linear-barrier solution agrees at the approximately $0.5$ per cent level for the 400-point variance grids used in the test suite. For a constant barrier, the moving-barrier term in the differentiated first-crossing equation of \citet{ZhangHui2006} vanishes identically, so the flat-barrier agreement is an identity of the implemented equation rather than an independent test of the numerical kernel; the linear-barrier and Monte Carlo comparisons are the tests of the kernel. The linear-barrier error is set by the grid spacing relative to the squared barrier shift and decreases as the grid is refined. The remaining discrepancy is controlled by the variance-grid discretisation rather than by an approximation to the underlying stochastic model.

\subsection{Direct Monte Carlo first-crossing cross-check}

For a genuinely nonlinear barrier, a closed-form first-crossing solution is not available. We therefore compare the numerical Volterra solution against direct Monte Carlo realisations of the underlying Markov random walks, reproducing the type of cross-check presented by \citet{ZhangHui2006}. The two calculations share the same barrier but reach the first-crossing distribution through independent numerical routes: one solves the probability conservation equation, while the other explicitly samples random trajectories and records their first barrier crossings. Agreement between them tests the first-crossing discretisation in a regime where the flat- and linear-barrier analytic limits do not apply. The Monte Carlo walks have independent increments, so this check validates the solver for the Markov problem and does not test the correlated top-hat case.

For the numerical first-crossing implementation, convergence must also be checked with respect to the variance grid and the integration range. The constant-barrier vectorised and compiled implementations use the analytic solution and are independent of this grid, whereas the serial implementation evaluates the numerical first-crossing equation. The validation tests therefore distinguish first-crossing discretisation from tree-timestep convergence.

\subsection{PCH08 branching-rate checks}
\label{sec:fortran_check}

\begin{figure}[t]
\centering
\includegraphics[width=\linewidth]{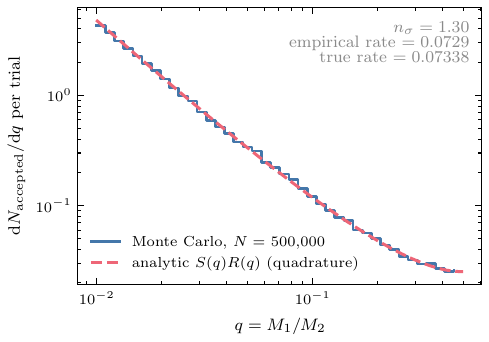}
\caption{PCH08 sampled progenitor mass ratio distribution (histogram; $n_{\rm trials}=5\times10^5$) compared with an independently coded quadrature of the target density (curve), for a step from $M_0=10^{12}\,\Msun/h$ at $z_0=0$ to $z_1=0.02$, with $M_{\rm res}=10^{10}\,\Msun/h$, on a Planck 2018 CAMB cosmology. Generated by \texttt{scripts/make\_validation\_figure.py}.}
\label{fig:branching_rate_validation}
\end{figure}

We test the PCH08 backend in two ways. First, we compare the sampled progenitor mass-ratio distribution with numerical quadrature of the density from which the sampler draws (Figure~\ref{fig:branching_rate_validation}). The two agree across $q=0.01$--$0.5$ (approximately $1.7$ decades). Since both calculations use the same implemented rate, this tests the sampling procedure rather than the rate itself.

Second, we compare the branching-rate calculation with the reference FORTRAN implementation of \citet{ParkinsonColeHelly2008}, which we compiled and ran. We supply \texttt{PCHMergerTree} with the same tabulated $\sigma(M)$, $\delta_{\rm c}(z)$ and $\alpha(M)$ as the FORTRAN code, together with the Millennium-like cosmology ($\Omega_{\rm m}=0.25$, $h=0.73$, $\sigma_8=0.9$, BBKS transfer function; $G_0=0.57$, $\gamma_1=0.38$, $\gamma_2=-0.01$). The single-step $N_{\rm upper}$ agrees with the FORTRAN expression to machine precision, while full-population branch counts agree with 300 realisations of the reference code to within $1$--$2$ per cent out to $z=5$. The comparison also caught a sign error in the PCH08 branching-rate normalisation during development that was not detected by the internal self-consistency checks.

We check the calculation of $\alpha$ separately for both the top-hat and sharp-$k$ window functions, comparing the implemented $\alpha=|\mathrm{d}\ln\sigma/\mathrm{d}\ln M|$ with a finite-difference derivative of the tabulated $\sigma(M)$.

\subsection{Cross-implementation comparison: PCH08 and Zhang--Hui trees}
\label{sec:pch08_vs_zh}

Both tree-generation strategies in \S~\ref{sec:architecture} use the same $S(M)$ and $B(S,z)$ for CDM. We can therefore compare them at matched initial conditions and resolution in a Planck cosmology. Any difference between the resulting trees then arises from the tree-generation prescription rather than from differences in the underlying mass variance, barrier or resolution.

We consider three aspects of this comparison: timestep compliance with the practical single-split criterion (\S~\ref{sec:tree_timestep}), numerical convergence as $\Delta z$ decreases, and the origin of any residual difference. We do not compare with $N$-body merger trees here, so these tests do not address which prescription more closely reproduces simulated haloes. The PCH08 calibration is likewise taken from \citet{ParkinsonColeHelly2008}.

\paragraph{Timestep compliance}
We first examine the dependence of the tree statistics on timestep and mass resolution (Table~\ref{tab:compliance}). For $M_0=10^{12}\,M_\odot/h$, reducing $M_{\rm res}/M_0$ from $10^{-2}$ to $10^{-4}$ causes a fixed $\Delta z=0.05$ step to violate the single-split criterion: the maximum PCH08 $N_{\rm upper}$ increases from $0.32$ to $14.4$, while the EPS expected split count for Zhang--Hui reaches $E\simeq7.5$. The resulting main-progenitor statistics change substantially as the timestep is reduced. At $M_{\rm res}/M_0=10^{-4}$, the PCH08 median $M(z=1)/M_0$ changes from $0.81$ at $\Delta z=0.05$ to $0.56$ at $\Delta z=0.002$, with no further change at $\Delta z=5\times10^{-4}$; the Zhang--Hui median changes from $0.49$ to $0.44$. The converged values therefore require substantially smaller timesteps at this resolution. 

\begin{table}[t]
\caption{Median main-progenitor mass at $z=1$, $M(z=1)/M_0$, for $M_0=10^{12}\,M_\odot/h$, from $3\,000$ trees per entry (sampling error $\approx0.01$; the compiled builders are not bitwise reproducible). $N_{\rm upper}$ is the largest mean-field PCH08 value along the trajectory. Square brackets give the PCH08 16--84 per cent range. Generated by \texttt{scripts/timestep\_compliance\_table.py}.}
\label{tab:compliance}
\centering
\footnotesize
\setlength{\tabcolsep}{4pt}
\begin{tabular}{@{}ccccc@{}}
\hline
$M_{\rm res}/M_0$ & $\Delta z$ & max $N_{\rm upper}$ & PCH08 & Zhang--Hui \\
 & & & median [16--84] & median \\
\hline
$10^{-2}$ & $0.05$ & $0.32$ & $0.55\ [0.36,\,0.68]$ & $0.45$ \\
$10^{-2}$ & $5\times10^{-4}$ & $0.00$ & $0.56\ [0.37,\,0.69]$ & $0.42$ \\
$10^{-4}$ & $0.05$ & $14.4$ & $0.81\ [0.79,\,0.81]$ & $0.49$ \\
$10^{-4}$ & $0.01$ & $2.8$ & $0.76\ [0.59,\,0.79]$ & $0.46$ \\
$10^{-4}$ & $0.002$ & $0.57$ & $0.56\ [0.36,\,0.69]$ & $0.45$ \\
$10^{-4}$ & $5\times10^{-4}$ & $0.14$ & $0.56\ [0.37,\,0.69]$ & $0.44$ \\
\hline
\end{tabular}
\end{table}

\begin{figure}[t]
\centering
\includegraphics[width=\linewidth]{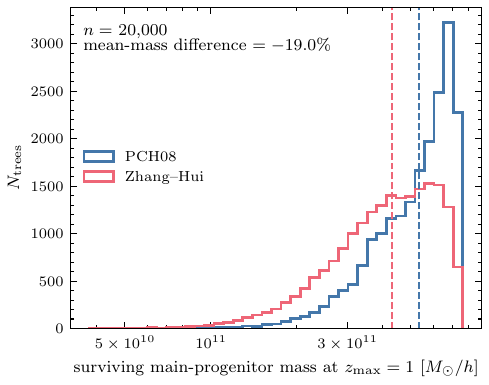}
\caption{Surviving main-progenitor mass distributions from the PCH08 and Zhang--Hui backends at matched $(M_0,z_0,z_{\rm max},M_{\rm res},\Delta z)=(10^{12}\,M_\odot/h,\,0,\,1,\,10^{10}\,M_\odot/h,\,0.005)$ with $n_{\rm trees}=20\,000$ on a Planck 2018 CAMB cosmology, showing the $\sim 19$ per cent offset in the mean surviving mass (dashed lines) of Section~\ref{sec:pch08_vs_zh}, where Table~\ref{tab:compliance} instead compares medians; $N_{\rm upper}<0.1$ at this $\Delta z$. Generated with \texttt{scripts/validate\_zhang\_hui\_vs\_pch08.py --output-figure}.}
\label{fig:pch08_vs_zhanghui}
\end{figure}

\paragraph{Residual difference at small timesteps}
A systematic difference remains once the timestep is sufficiently small. Table~\ref{tab:compliance} compares the median surviving mass: for $M_{\rm res}/M_0=10^{-2}$, the PCH08 median $M(z=1)/M_0$ is $0.56$, compared with $0.42$ for Zhang--Hui; at $M_{\rm res}/M_0=10^{-4}$ the corresponding values are $0.56$ and $0.44$. The Zhang--Hui median is therefore lower by $\sim21$--$25$ per cent. Figure~\ref{fig:pch08_vs_zhanghui} shows the full surviving-mass distributions for $M_{\rm res}/M_0=10^{-2}$ and $\Delta z=0.005$; the mean surviving mass is lower for Zhang--Hui by $\sim19$ per cent. The precise offset depends on the statistic, but its persistence as $\Delta z$ decreases shows that it is not a timestep effect.

The difference arises from the different progenitor prescriptions used by the two algorithms. PCH08 uses its fitted, N-body-calibrated branching-rate prescription together with the rejection correction, whereas Zhang--Hui uses the closed-form first-crossing distribution for the specified constant-barrier Markov problem. These different approximations to the conditional progenitor mass function produce different resolved merger-mass loss.

We can reproduce this difference without generating merger trees. Starting from the same initial halo, we propagate the mean progenitor mass using the transition probabilities used by each algorithm. This gives a difference of $-18.9$ per cent, compared with $-19.0$ per cent from an independent 20,000-tree Monte Carlo calculation. The agreement to $0.1$ percentage points shows that the difference in the mean surviving mass is already present in the transition probabilities themselves.

\begin{figure*}[t]
\centering
\includegraphics[width=\linewidth]{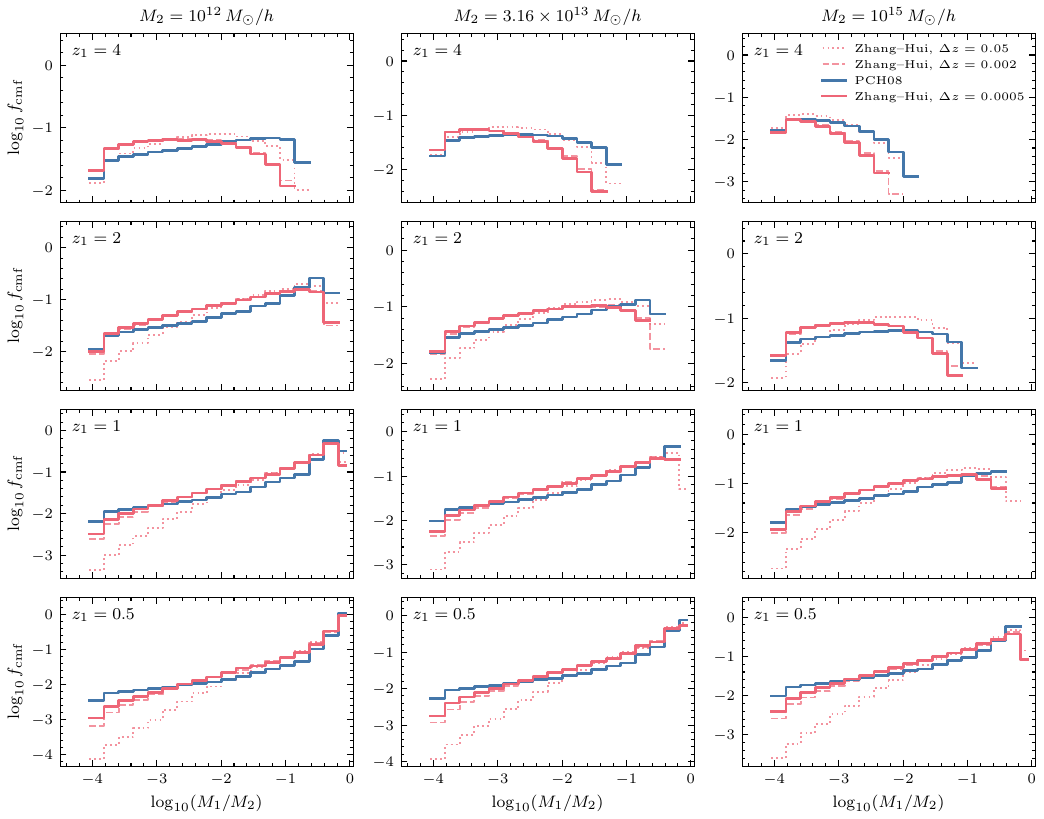}
\caption{Mass-weighted conditional mass function $f_{\rm cmf}(M_1|M_2)$ for $M_2\in\{10^{12},3.16\times10^{13},10^{15}\}\,M_\odot/h$ (columns) and $z_1\in\{0.5,1,2,4\}$ (rows), $M_{\rm res}/M_2=10^{-4}$, $n_{\rm trees}=1\,000$. PCH08 (blue) uses adaptive $N_{\rm upper}$-compliant stepping (\texttt{--target-nupper 0.1}, ceiling $\Delta z=0.05$). Zhang--Hui (red) uses a fixed step: $\Delta z=5\times10^{-4}$ (solid), $0.002$ (dashed) and $0.05$ (dotted). 
Generated with \texttt{scripts/paper\_figs/fig1\_conditional\_mass\_function.py}.}
\label{fig:pch08_vs_zh_cmf}
\end{figure*}

\begin{figure}[t]
\centering
\includegraphics[width=\linewidth]{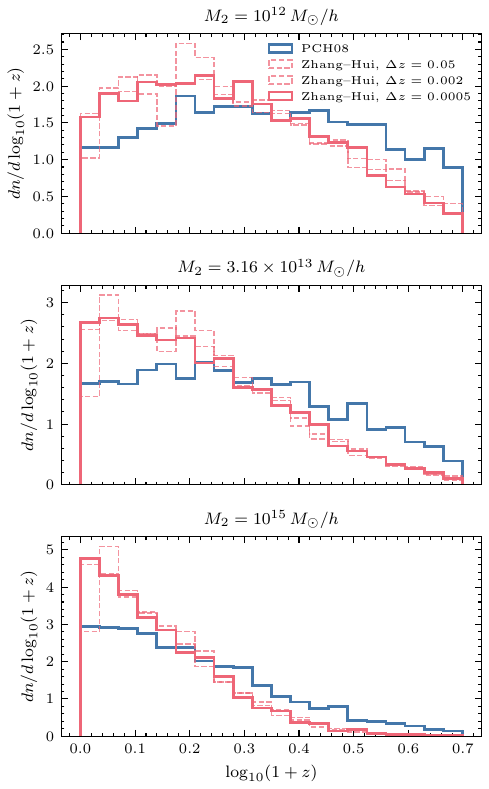}
\caption{Redshift of the main progenitor's most recent major merger (mass ratio $\geq 0.3$) for $M_2\in\{10^{12},3.16\times10^{13},10^{15}\}\,M_\odot/h$, $M_{\rm res}/M_2=10^{-4}$, $n_{\rm trees}=5\,000$. PCH08 (blue) uses adaptive $N_{\rm upper}$-compliant stepping (ceiling $\Delta z=0.05$). Zhang--Hui (red) uses a fixed step: $\Delta z=5\times10^{-4}$ (solid), $0.002$ and $0.05$ (thin dashed). Generated with \texttt{scripts/paper\_figs/fig3\_major\_merger\_redshift.py}.}
\label{fig:pch08_vs_zh_major_merger}
\end{figure}

\paragraph{Conditional mass function and major mergers}
Figures~\ref{fig:pch08_vs_zh_cmf} and~\ref{fig:pch08_vs_zh_major_merger} compare statistics accumulated over complete merger trees. Figure~\ref{fig:pch08_vs_zh_cmf} shows the mass-weighted conditional mass function $f_{\rm cmf}(M_1|M_2)$: the fraction of a descendant halo's mass at $z_2=0$ found in progenitors of mass $M_1$ at $z_1=0.5,1,2,4$, for three final masses spanning group to cluster scales. The comparison uses full multi-generation branch populations with $M_{\rm res}/M_2=10^{-4}$. PCH08 uses adaptive $N_{\rm upper}$-compliant steps, while Zhang--Hui is shown at $\Delta z=5\times10^{-4}$; the earlier $\Delta z=0.05$ calculation is retained as a non-compliant reference.

The Zhang--Hui result has substantial timestep dependence at $\Delta z=0.05$, particularly near the resolution limit. At $\Delta z=0.002$ it is already close to the $\Delta z=5\times10^{-4}$ result, with differences of at most $0.22\,{\rm dex}$. The $\Delta z=5\times10^{-4}$ result is therefore sufficiently converged for the comparison. At this timestep, the two backends differ by up to $\sim0.4\,{\rm dex}$ over most of the resolved mass range. Away from the resolution and high-redshift extremes, PCH08 is typically lower than Zhang--Hui by $0.06$--$0.17\,{\rm dex}$. The difference grows towards the ends of the resolved range, reaching $\sim0.5\,{\rm dex}$ at low mass ratios and up to $\sim0.8\,{\rm dex}$ at the highest mass ratios at $z_1=4$.

Figure~\ref{fig:pch08_vs_zh_major_merger} shows the redshift of the most recent major merger on the main branch, defined by a mass ratio $\geq0.3$. The Zhang--Hui median is stable to timestep refinement: for $M_2=10^{12}\,M_\odot/h$ it changes from $0.85$ at $\Delta z=0.05$ to $0.80$ and $0.82$ at $\Delta z=0.002$ and $5\times10^{-4}$, respectively. At $\Delta z=5\times10^{-4}$, the PCH08 medians are $1.14$, $0.88$ and $0.52$ for $M_2=10^{12}$, $3.16\times10^{13}$ and $10^{15}\,M_\odot/h$, compared with $0.82$, $0.56$ and $0.32$ for Zhang--Hui. Major mergers occur in $87$, $94$ and $99$ per cent of PCH08 trees, compared with $97$, $99$ and $99.9$ per cent of Zhang--Hui trees.


\subsection{Constrained-tree validation}
\label{sec:constrained_validation}

We validate the constrained implementation at three levels.

First, the constrained integral-equation solution approaches the corresponding unconstrained solution as $S_1\rightarrow\infty$, as expected from the limiting behaviour discussed by \citet{NadlerEtAl2023}. With matched discretisation, the two solutions agree to better than $1$ per cent once the constraint is sufficiently remote.

Second, we compare the sample-path Brownian-bridge calculation used in tree construction with two independent calculations of the constrained first-crossing distribution for a flat barrier. The first solves the constrained integral equation directly; the second generates unconstrained random walks and retains only trajectories entering a narrow window around $(S_1, \delta_1)$. The resulting cumulative first-crossing probabilities agree to within about 2 per cent for the rejection-sampling calculation, and to within about 3–4 per cent for the sample path sampler at the step size used in the test ($dS/S_1$ = 0.004). The sample-path offset is a systematic one, since crossings between grid points are missed, and it decreases as $dS$ is reduced.

As a further check, we compared these results with an analytic expression for the first-crossing distribution of a Brownian bridge with a linear barrier, $B(S)=b_0+b_1S$, ending on the barrier. We obtained this from the first-crossing distribution of the unconstrained random walk by requiring that walk end at the constraint point. The probability of first crossing in each grid cell agrees with the analytic value to a relative precision of $\sim10^{-11}$ for flat and linear barriers. Note that this is a test of the integral-equation solver only; the tree-level sampler only supports flat barriers and approaches this solution only as d$S\rightarrow0$ because it samples the bridge on a grid of step d$S$.

\begin{figure}[t]
\centering
\includegraphics[width=\linewidth]{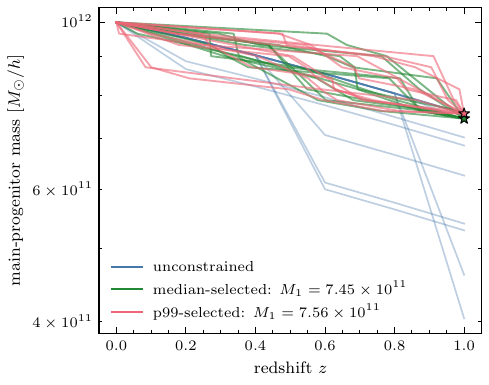}
\caption{Main-branch growth histories from the Brownian-bridge-constrained sampler for $(M_0,z_0,z_1)=(10^{12}\,\Msun/h,0,1)$ and $M_{\rm res}=10^{10}\,\Msun/h$, with $n_{\rm trees}=3000$. The two constraint masses are the median ($7.45\times10^{11}\,\Msun/h$) and 99th percentile ($7.56\times10^{11}\,\Msun/h$) of the unconstrained masses at $z=1$. 
Generated by \texttt{scripts/validate\_constrained\_tree\_convergence.py}.}
\label{fig:constrained_vs_unconstrained}
\end{figure}

Finally, we test the constrained main-branch sampler end to end. For $(M_0,z_0,z_1)=(10^{12}\,M_\odot/h,0,1)$ and $M_{\rm res}=10^{10}\,M_\odot/h$, we first generate 3000 unconstrained histories with the general Zhang--Hui builder ($\Delta z=0.2$, $N_{\rm grid}=60$). From their $z=1$ mass distribution we choose two constraint masses: the median, $M_1=7.45\times10^{11}\,M_\odot/h$, and the 99th percentile, $M_1=7.56\times10^{11}\,M_\odot/h$. We then generate 3000 constrained histories for each constraint using a bridge step of $\mathrm{d}S=0.05$. Every constrained history reaches the specified $M_1$ at $z_1$, and the mean masses at $z=0.5$ are $1.079\pm0.003$ and $1.085\pm0.003$ times the corresponding unconstrained mean. The larger constraint gives $0.0049\pm0.0012\,M_0$ more mass at $z=0.5$, as expected (Figure~\ref{fig:constrained_vs_unconstrained}).

The reference histories in this test are deliberately generated with $\Delta z=0.2$ and $N_{\rm grid}=60$, rather than with a converged calculation: $\Delta z=0.2$ does not satisfy the single-split criterion and the variance grid under-resolves the split probability. The two constraint masses therefore define test cases for the constrained sampler rather than converged predictions for the halo population.

\section{Performance}
\label{sec:performance}

The computational implementations described in \S~\ref{sec:architecture} provide different trade-offs between simplicity and computational efficiency. \foraois implements a serial backend, a vectorised backend that evolves an ensemble of trees together, and a just-in-time (JIT) compiled backend that additionally exploits thread-level parallelism across independent trees.

\begin{figure}[t]
\centering
\includegraphics[width=\linewidth]{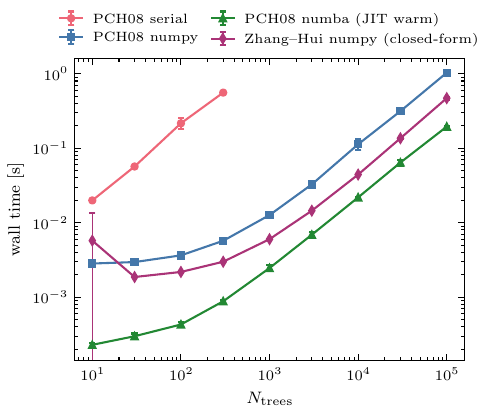}
\caption{Wall time versus forest size $N$ for the PCH08 serial, vectorised (NumPy), and compiled (Numba; JIT-warm) implementations, together with the unconstrained Zhang--Hui backend's vectorised (closed-form) implementation. The benchmark uses fixed $(z_0,z_{\rm max},\Delta z)=(0,5,0.2)$ and $M_{\rm res}=10^9\,\Msun/h$ with a synthetic power-law $P(k)$. Points show the mean and standard deviation of five repeats. These settings measure computational scaling at a fixed $\Delta z$.
}
\label{fig:backend_benchmark}
\end{figure}

Figure \ref{fig:backend_benchmark} shows the results from the benchmark tests. All timings reported in this section were measured with \texttt{scripts/benchmark\_backends.py --repeats 5}, which re-times each $(N,\text{backend})$ point five times within a single warm process (Numba JIT compilation is a one-time cost which we exclude here) and reports the mean and standard deviation. The benchmark runs were performed on an 8-core Apple M1 (arm64) laptop running Python 3.11, NumPy 2.4, SciPy 1.17, and Numba 0.67, and a synthetic power-law $P(k)$ is used throughout (no CLASS/CAMB dependency).

On this machine, the compiled (numba) PCH08 backend is consistently faster than the vectorised (numpy) implementation across the full range of $N$ tested, by a factor of $\sim 5.3$ at the largest forest size benchmarked ($N=100\,000$: $1.034\pm0.004\,\mathrm{s}$ numpy versus $0.195\pm0.003\,\mathrm{s}$ numba), and both are substantially faster per tree than the serial backend. We caution that these numbers provide an illustrative performance comparison: the compiled backend's thread-parallel speed-up over the vectorised backend is bounded by the number of available cores, and the precise speed-up factor is machine- and compiler-dependent. These benchmarks characterise single-node, shared-memory performance on one 8-core machine; scaling to larger shared-memory or distributed systems is outside the scope of this work.

The parallel Numba backend does not provide bitwise reproducibility of individual realisations from a fixed seed, although ensemble statistics remain reproducible.

The practical implication is that standard constant-barrier calculations can exploit analytic and compiled fast paths to generate large ensembles efficiently. The general numerical first-crossing solver, which a moving barrier requires, is more expensive than either closed-form path. It is demonstrated at the first-crossing level and is not benchmarked here, and tree-level moving-barrier calculations are not demonstrated in this paper.

Figure~\ref{fig:backend_benchmark} also includes the closed-form Zhang--Hui backend's vectorised (NumPy) implementation, timed under the same conditions. At $N=100\,000$ and $\Delta z=0.2$, it takes $0.471\pm0.022\,\mathrm{s}$ against $1.034\pm0.004\,\mathrm{s}$ for the PCH08 NumPy implementation. Because that comparison sits at a step that is not compliant for either algorithm, Table~\ref{tab:benchmark} compares both algorithms, in both implementations, at matched $\Delta z$ down to a compliant step, on a Planck 2018 CAMB cosmology for $M_0=10^{12}\,M_\odot/h$, $M_{\rm res}=10^{9}\,M_\odot/h$ and $z=0$--$5$.

\begin{table*}[t]
\caption{Wall time in seconds for $N=10\,000$ main-progenitor histories, $M_0=10^{12}\,M_\odot/h$, $M_{\rm res}=10^{9}\,M_\odot/h$, $z=0$--$5$, on the machine described above (mean of three repeats after warm-up; CAMB power spectrum). The largest mean-field PCH08 $N_{\rm upper}$ and the largest EPS expected number of splits per step $E$ along the trajectory indicate which steps meet the requirement of $\lesssim0.1$. Generated by \texttt{scripts/benchmark\_matched\_dz.py}.}
\label{tab:benchmark}
\centering
\small
\begin{tabular}{ccccccc}
\hline
$\Delta z$ & max $N_{\rm upper}$ & max $E$ & PCH08 NumPy & PCH08 Numba & Zhang--Hui NumPy & Zhang--Hui Numba \\
\hline
$0.2$ & $8.2$ & $7.6$ & $0.11$ & $0.03$ & $0.06$ & $0.02$ \\
$0.01$ & $0.39$ & $0.38$ & $2.36$ & $0.33$ & $1.09$ & $0.22$ \\
$0.003$ & $0.12$ & $0.12$ & $8.19$ & $1.01$ & $3.80$ & $0.67$ \\
$0.001$ & $0.039$ & $0.038$ & $25.0$ & $2.72$ & $11.7$ & $1.84$ \\
\hline
\end{tabular}
\end{table*}

Both algorithms scale approximately as $1/\Delta z$ in both implementations. At matched $\Delta z$ the closed-form Zhang--Hui builder is about $1.8$ times cheaper than PCH08 in NumPy at $\Delta z=0.2$, rising to $2.1$--$2.2$ at smaller $\Delta z$, and about $1.5$ times cheaper in Numba across the range tested. The timestep dominates the cost: meeting the criterion here needs $\Delta z\approx0.003$ or smaller, roughly $30$ to $100$ times the cost of $\Delta z=0.2$ for either algorithm. 

\section{Example application: \textit{Ashvini}}
\label{sec:ashvini}

As an example of how \texttt{foraois} interfaces with other codebases, we use it to provide halo merger histories to \textit{Ashvini} \citep{MenonPower2024,MenonBaluPower2026}, a lightweight semi-analytic model of galaxy formation. \textit{Ashvini} integrates star formation, supernova and AGN feedback, and reionisation suppression along a halo's merger history. We use the model to calculate an illustrative $z=0$ stellar-to-halo mass relation (SHMR) for CDM, $3\,{\rm keV}$ thermal-relic WDM and $10^{-22}\,{\rm eV}$ FDM. In the present implementation, WDM and FDM enter \texttt{foraois} through the linear matter power spectrum alone (\S\ref{sec:dm_models}): all three models use the same real-space top-hat window and a mass-independent collapse barrier, and the barriers are not independently calibrated for WDM or FDM. The top-hat treatment is therefore not intended to provide a quantitatively calibrated prediction of low-mass WDM or FDM halo suppression; the comparison is instead an illustrative demonstration of software interoperability.

For each model and each of 13 logarithmically spaced halo masses between $10^8$ and $10^{14} M_\odot$ at $z=0$, we generate 1000 Zhang--Hui main-progenitor histories with $\Delta z=0.005$, $M_{\rm res}=10^{-4}M_0$, and $z_{\rm max}=30$, and integrate the baryonic model along each history. The halo growth rate supplied to \textit{Ashvini} is constructed from the smooth-accretion and merger-growth components returned by the tree. These runs use a fixed redshift grid chosen to be consistent with the galaxy-formation calculation, rather than a timestep selected to satisfy the merger-tree convergence criterion.

This calculation is for demonstrating interoperability, not physical fidelity. At $z=0$, the expected EPS number of splits per timestep is approximately $0.1$ at $10^9 M_\odot$ and $0.7$ at $10^{12} M_\odot$, so the adopted timestep exceeds the $\lesssim0.1$ criterion used for PCH08 timestep control at the high-mass end. The resulting stellar masses should consequently not be interpreted as numerically converged with respect to the merger-tree timestep or mass resolution.

\subsection{Coupling to \textit{Ashvini}}
The coupling requires only a small adapter in \textit{Ashvini}. For each halo ensemble, the adapter constructs a \texttt{ZhangHuiMergerTree} from the \texttt{foraois} cosmological data and run parameters and calls \texttt{build\_forest\_numba}. The resulting merger histories provide the halo mass at each timestep together with the smooth-accretion and merger-mass components. The adapter combines these components to form the total halo growth rate.

The sole input to \textit{Ashvini} is the merger histories, i.e. the halo mass, halo growth rate and redshift. The baryonic prescriptions, including star formation, feedback and reionisation, are therefore identical for the three dark matter models. The three calculations differ only in the \texttt{foraois} configuration specifying the dark matter model and its associated mass scale; the tree algorithm, random seed, timestep, mass resolution, maximum redshift, number of realisations and \texttt{pk\_kmax} override are otherwise identical. Here \texttt{pk\_kmax} sets the upper wavenumber cutoff in the integral used to calculate $\sigma(M)$ and its logarithmic derivative $\alpha(M)$ from the linear matter power spectrum. Holding it fixed ensures that differences between the three calculations are not introduced by different numerical cutoffs in the power-spectrum integration.

\subsection{Illustrative SHMR calculation}
\begin{figure*}[t] 
\centering 
\includegraphics[width=\linewidth]{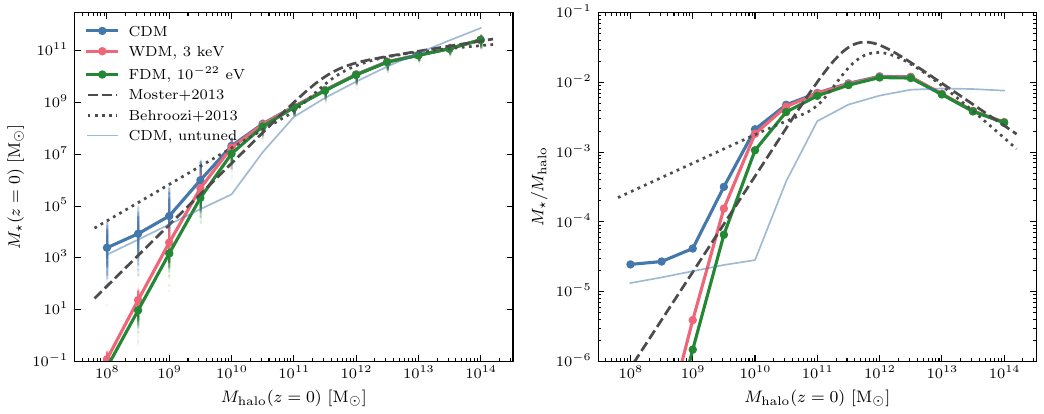} 
\caption{Median $z=0$ stellar mass (left) and $M_\star/M_{\rm halo}$ (right) from \textit{Ashvini} run on \texttt{foraois} Zhang--Hui trees, for CDM, $3\,{\rm keV}$ thermal-relic WDM and $10^{-22}\,{\rm eV}$ FDM (1000 realisations per halo-mass bin
). The baryonic model is the illustrative configuration described in Section~\ref{sec:ashvini}. WDM and FDM enter through the linear power spectrum alone, with an uncalibrated mass-independent barrier and a top-hat window, so the comparison is an illustrative demonstration, not a calibrated prediction. Dashed and dotted lines are the abundance-matching fits of \citet{MosterNaabWhite2013} and \citet{BehrooziWechslerConroy2013}; the thin blue line is CDM with \textit{Ashvini}'s default parameters.} \label{fig:ashvini_shmr_dm_models} 
\end{figure*}

For this demonstration (Figure~\ref{fig:ashvini_shmr_dm_models}) we use an illustrative \textit{Ashvini} configuration with $\epsilon_{\rm ff}=0.05$ and a fixed hot-mode prescription with $M_{\rm hot}=2\times10^{12}\,M_\odot$, $f=0.1$ and $\phi=4$; the reionisation suppression parameter $\omega=2$ is \textit{Ashvini}'s default value, unchanged here. The three tuned values differ from the \textit{Ashvini} defaults and were selected to bring the CDM relation closer to the abundance-matching relation of \citet{BehrooziWechslerConroy2013}. The resulting model is illustrative rather than a physical calibration.

The tuned CDM relation is within approximately $0.2$--$0.25\,{\rm dex}$ rms of the \citet{BehrooziWechslerConroy2013} relation over the halo-mass ranges shown, although it remains below that relation near the peak of the SHMR. This provides a check that the end-to-end calculation produces stellar masses of the expected order of magnitude.

The dark matter models produce similar stellar masses at high halo mass and progressively different results towards the low-mass end. In the regenerated tuned calculation, the WDM and FDM median stellar masses relative to CDM are approximately $0.82$ and $0.49$ at $10^{10}\,M_\odot$, and $0.09$ and $0.04$ at $10^9\,M_\odot$. The magnitude of these low-mass differences depends on the galaxy-formation prescriptions, particularly the treatment of the UV background, and on the uncalibrated WDM/FDM collapse treatment noted above.

This example demonstrates the intended separation between halo-tree generation and galaxy-formation modelling: the dark-matter dependence enters through the halo histories rather than through the galaxy-formation code, with \texttt{foraois} supplying halo mass, growth rate and redshift to an unmodified \textit{Ashvini} calculation.

\section{Discussion}
\label{sec:discussion}

The design of \texttt{foraois} keeps the linear power spectrum, smoothing and mass assignment, collapse barrier, first-crossing calculation, and numerical tree construction separate. This makes it possible to change the cosmology or collapse model without changing the tree-building algorithm, and to identify which differences between merger trees arise from each component.

PCH08 is calibrated to simulated merger trees, whereas the Zhang--Hui construction is not. The two therefore make different approximations to the progenitor distribution. There are also numerical differences between the tree builders: both allow at most one split per step, while the Zhang--Hui binary draw approximates the EPS tree. We therefore compare the two at timesteps satisfying the single-split criterion and check the results as $\Delta z$ is reduced. A residual difference remains at such timesteps (\S~\ref{sec:pch08_vs_zh}). The transition probabilities used by the two algorithms reproduce essentially the full difference in the tested configuration, showing that it arises from their different progenitor prescriptions. This leaves the Markov approximation used to construct the top-hat $S(M)$ (\S~\ref{sec:variance_barrier}) as a separate source of uncertainty, and comparison with $N$-body merger trees is needed to determine how either prescription fares against simulated haloes.

The computational cost depends on both the first-crossing calculation and the number of progenitors being evolved. Analytic constant-barrier main-branch ensembles are inexpensive, whereas repeatedly solving for every resolved progenitor is substantially more costly. At matched step size the closed-form Zhang--Hui builder in \texttt{foraois} is cheaper than the PCH08 builder (\S~\ref{sec:performance}), but the timestep required to satisfy the single-split criterion dominates the cost of both, as shown in Table~\ref{tab:benchmark}. The two backends therefore provide different trade-offs rather than a single preferred tree-building method.

The constrained implementation currently accepts one conditioning point $(M_1,z_1)$. Brownian bridges could be extended to multiple conditioning points, or to constraints on derived observables such as a galaxy property related to halo mass through a semi-analytic or abundance-matching model. These extensions are not implemented here. Secondary subtrees are also not grown along the constrained branch, so the merger mass there is set to zero (\S~\ref{sec:accretion_decomposition}); the constrained histories therefore describe main-branch growth along the constrained segment, with growth assigned to the smooth-accretion channel.

For WDM, the constant collapse barrier used here is calibrated for CDM. It remains to determine whether this barrier is adequate for quantitative work in WDM, or whether the branching-rate correction should also be refitted for WDM, as in \citet{BensonEtAl2013} for PCH08. This is the main unresolved physical question in the present WDM implementation.

Finally, the single-split criterion provides a practical limit on the timestep for both tree builders. At fixed $\Delta z$, PCH08 main-progenitor histories become nearly deterministic and assemble too early when $N_{\rm upper}\gg1$, which occurs as $M_{\rm res}/M_0$ decreases (Table~\ref{tab:compliance}). Zhang--Hui has the corresponding limitation through the EPS expected number of splits per step. Users should therefore check the relevant quantity in \texttt{foraois.diagnostics} for their chosen $(M_0,M_{\rm res},\Delta z)$.

\section{Summary and conclusions}
\label{sec:summary}

We have presented \texttt{foraois}, a Python framework designed to separate the distinct modelling assumptions that enter Monte Carlo dark-matter halo merger histories. The contributions of cosmology, collapse physics, first-crossing statistics, and tree construction are treated separately: the linear matter power spectrum determines the variance $S(M)$; the collapse prescription determines the barrier $B(S,z)$; a first-crossing calculation determines the corresponding progenitor statistics; and a tree-building algorithm samples those statistics to construct discrete halo histories. Keeping these stages separate allows each part of the calculation to be varied and tested independently.

The implementation includes both the \citet{ParkinsonColeHelly2008} (PCH08) and \citet{ZhangHui2006} approaches for merger-tree construction, together with a numerical first-crossing solver that can treat barriers beyond the constant-barrier case. The Brownian-bridge machinery extends this to constrained halo assembly histories. The resulting components can be combined so that the tree-building algorithm is independent of how the power spectrum or collapse barrier was obtained.

We have tested the different stages of the calculation separately. The first-crossing solver has been compared with analytic and independently calculated solutions, the Monte Carlo random-walk calculation has been checked against the corresponding Volterra solution, and the PCH08 branching calculation has been compared with the published FORTRAN implementation. We have also compared the PCH08 and Zhang--Hui merger trees and tested the constrained-history calculation independently. These tests probe different parts of the calculation: agreement in the first-crossing solution, for example, does not by itself imply agreement in the statistics of a discrete merger-tree sampler.

The comparison of the two tree-building algorithms illustrates why this separation is useful. When the timestep satisfies the single-split criterion, the PCH08 and Zhang--Hui trees retain a systematic difference even as the timestep is reduced. We can reproduce essentially the full difference in mean surviving mass from the different transition probabilities used by the two algorithms. The difference is therefore a property of their progenitor prescriptions, rather than a residual timestep effect. At the same time, the two algorithms can be applied to the same cosmological inputs and collapse barrier, allowing these differences to be studied without changing the underlying halo model. This provides a direct way to separate differences arising from the progenitor prescription from those arising from the cosmology or collapse model.

The coupling to \texttt{Ashvini} \citep{MenonBaluPower2026} provides a second example of this separation. Halo histories generated by \texttt{foraois} can be passed to a galaxy-formation calculation without changing its physical model. More generally, \texttt{foraois} can be used with histories generated from different dark-matter power spectra, currently CDM, WDM, and FDM, or with different collapse prescriptions through the serial Zhang--Hui builder. This makes it possible to change the halo model while holding the subsequent galaxy-formation calculation fixed, and conversely to study the response of the galaxy model to different halo histories without rebuilding the halo calculation.

There are also important limits to the physical conclusions that can be drawn from the present implementation. The first-crossing solver assumes Markovian random walks, whereas the default top-hat smoothing produces correlated steps; the resulting first-crossing calculation is therefore an approximation whose impact has not been quantified here. The WDM and FDM calculations likewise use a constant collapse barrier calibrated for CDM. Whether a modified barrier or branching correction is required for quantitative predictions in these models remains to be established. Finally, the PCH08 and Zhang--Hui constructions produce systematically different merger histories even when the timestep is sufficiently small. We have shown that this difference arises from their different progenitor prescriptions, but comparison with $N$-body merger trees is needed to determine which prescription more accurately represents halo assembly.

These limitations also define the scientific use of \texttt{foraois}. Its purpose is not to hide the approximations involved in constructing merger trees, but to expose them. Cosmological inputs, collapse prescriptions, first-crossing calculations, and tree-building algorithms can be varied independently and their effects on halo assembly histories measured separately. The PCH08--Zhang--Hui comparison provides one example: using the same cosmology and collapse barrier, we can identify a systematic difference between the resulting trees and trace it to the progenitor prescriptions of the two algorithms. The same separation allows halo histories generated under different dark-matter models or collapse assumptions to be passed to a fixed galaxy-formation calculation. \texttt{foraois} therefore provides a way to investigate how assumptions about the dark-matter density field, halo collapse, and merger-tree construction propagate into the halo histories on which models of galaxy formation depend.


\section{Availability}
\label{sec:availability}

\texttt{foraois} is developed openly at \url{https://github.com/doctorcbpower/foraois} under the MIT licence. The implementation described here is release \texttt{v0.1.2}, archived at \url{https://doi.org/10.5281/zenodo.23093138}. We recommend using this release or later commits.

The example in Section~\ref{sec:ashvini} was produced with \textit{Ashvini} (\texttt{scripts/shmr\_dm\_model\_comparison.py --variant tuned}, repository \url{https://github.com/doctorcbpower/Ashvini}). The other figures and tables were produced with \texttt{v0.1.2}; \texttt{docs/PAPER\_FIGURES.md} lists the script and command for each, and the data behind the figures made from Monte Carlo runs are in \texttt{figures/data}. The tabulated reference data used for the FORTRAN comparison are included in the repository.

The package is not yet distributed through PyPI and can be installed directly from the repository:
\begin{lstlisting}
git clone git@github.com:doctorcbpower/foraois.git
cd foraois
pip install -e .
\end{lstlisting}
This installs the core dependency set described in \S~\ref{sec:usage_extras}. Optional CAMB, CLASS, and Numba functionality can be installed using, for example,
\begin{lstlisting}
pip install -e '.[camb]'
pip install -e '.[class]'
pip install -e '.[numba]'
\end{lstlisting}

The example configuration files, validation and benchmark scripts, and test suite referenced throughout this paper are part of the repository checkout rather than the installed runtime package. Continuous integration tests supported Python versions on Linux and macOS, builds the source and wheel distributions, and verifies that the resulting installation can be imported outside the repository checkout.

\section*{Acknowledgements}
We thank the referee warmly for their constructive report. This research uses \textsc{python} \citep{vanRossum.1995} libraries, including \textsc{numpy} \citep{Harris2020}, \textsc{matplotlib} \citep{Hunter2007}, and \textsc{SciencePlots} \citep{SciencePlots}.

\section*{Funding}
SB acknowledges the support of grant PID2022-138855NB-C32 funded by MICIU/AEI/10.13039/501100011033 and ERDF/EU, and  project PPIT2024-31833, cofunded by EU-Ministerio de Hacienda y Función P\'{u}blica-Fondos Europeos-Junta de Andaluc\'{i}a-Consejería de Universidad, Investigaci\'{o}n e Innovaci\'{o}n. 

\section*{CRediT authorship contribution statement}
\textbf{C. Power:}
Conceptualization, Methodology, Software, Validation, Writing -- original draft, Writing -- review \& editing.

\textbf{S. Balu:}
Methodology, Software, Validation, Writing -- review \& editing.

\section*{Declaration of competing interest}
The authors declare that they have no known competing financial interests or personal relationships that could have appeared to influence the work reported in this paper.

\section*{Declaration of generative AI and AI-assisted technologies in the manuscript preparation process}
During the preparation of this work the authors used Anthropic's Claude in order to optimise code and automate tests, and perform consistency checks on the manuscript. After using this tool/service, the authors reviewed and edited the content as needed and take full responsibility for the content of the published article.


\bibliographystyle{elsarticle-harv}
\bibliography{main}

\end{document}